\documentclass[12pt]{article}
\usepackage[top=1in, left = 1in, bottom = 1in, right = 1in, paperheight = 11in, paperwidth = 8.5in]{geometry}

\usepackage[utf8]{inputenc}
\usepackage{adjustbox}
\usepackage{sectsty}
\usepackage[hyphens]{url}
\usepackage[breaklinks]{hyperref}
\usepackage{amsmath, amssymb, amsthm, amsbsy, amsfonts, cases}
\usepackage{graphicx, comment}
\usepackage[noend]{algpseudocode}
\usepackage{bigints}
\usepackage[mathscr]{euscript}
\usepackage[makeroom]{cancel}
\usepackage{graphicx}
\usepackage[table]{xcolor}
\usepackage{subcaption}
\usepackage{enumitem}
\usepackage{verbatim}
\usepackage[font={small,it,singlespacing}]{caption}
\usepackage{lscape}
\usepackage{caption}
\usepackage{hyperref}
\usepackage[sort]{natbib}
\usepackage{parskip}
\usepackage{array}
\usepackage{multirow}
\usepackage{graphicx}
\usepackage{wrapfig}
\usepackage{lscape}
\usepackage{rotating}
\usepackage{epstopdf}
\usepackage[compact]{titlesec}
\usepackage{authblk}
\usepackage[toc,page]{appendix}
\usepackage{setspace}
\usepackage{xr}
\usepackage[sort]{natbib}
\usepackage{color, colortbl}
\definecolor{Gray}{gray}{0.9}
\usepackage{algorithm}
\usepackage{algpseudocode}
\usepackage{boldline}
\usepackage[letterspace=-9]{microtype}
\usepackage{ragged2e}

\hypersetup{colorlinks,citecolor=blue}
\graphicspath{{./images//}}

\setlist{nosep}

\sectionfont{\fontsize{12}{12}\selectfont}
\subsectionfont{\fontsize{12}{12}\selectfont}
\subsubsectionfont{\fontsize{12}{12}\selectfont}

\theoremstyle{plain}

\newtheoremstyle{exampstyle}
  {\topsep} 
  {\topsep} 
  {} 
  {} 
  {\bfseries} 
  {} 
  {.5em} 
  {} 

\newtheoremstyle{exampstyle}
  {\topsep} 
  {\topsep} 
  {} 
  {} 
  {\bfseries} 
  {.} 
  {.5em} 
  {} 

\providecommand{\keywords}[1]
{
  \small	
  \textbf{\textit{Keywords---}} #1
}

\usepackage{xr}
\makeatletter
\newcommand*{\addFileDependency}[1]{
  \typeout{(#1)}
  \@addtofilelist{#1}
  \IfFileExists{#1}{}{\typeout{No file #1.}}
}
\makeatother

\usepackage{algpseudocode}
\usepackage{algorithm}

\algnewcommand\algorithmicinput{\textbf{Input:}}
\algnewcommand\Input{\item[\algorithmicinput]}

\algnewcommand{\algorithmicoutput}{\textbf{return:}}
\algnewcommand\Output{\item[\algorithmicoutput]}

\algnewcommand\algorithmicforeach{\textbf{for each}}
\algdef{S}[FOR]{ForEach}[1]{\algorithmicforeach\ #1\ \algorithmicdo}

\theoremstyle{plain}

\title{\vspace{-0.5in}\normalsize{\textbf{But Can You Use It? Design Recommendations for Differentially Private Interactive Systems}}\vspace{-1em}}
  
\author[1]{\small{Liudas Panavas}}
\author[2]{\small{Joshua Snoke}}
\author[3]{\small{Erika Tyagi}}
\author[3]{\small{Claire McKay Bowen}}
\author[3]{\small{Aaron R. Williams}\vspace{-1em}}

\affil[1]{Northeastern University\\ panavas.l@northeastern.edu}
\affil[3]{Urban Institute\\ etyagi@urban.org \& cbowen@urban.org \& awilliams@urban.org}
\affil[2]{RAND Corporation\\ jsnoke@rand.org}

\date{}

\begin{document}
\doublespacing
\lsstyle

\maketitle

\vspace{-30pt}

\begin{abstract}
Accessing data collected by federal statistical agencies is essential for public policy research and improving evidence-based decision making, such as evaluating the effectiveness of social programs, understanding demographic shifts, or addressing public health challenges. Differentially private interactive systems, or validation servers, can form a crucial part of the data-sharing infrastructure. They may allow researchers to query targeted statistics, providing flexible, efficient access to specific insights, reducing the need for broad data releases and supporting timely, focused research. However, they have not yet been practically implemented. While substantial theoretical work has been conducted on the privacy and accuracy guarantees of differentially private mechanisms, prior efforts have not considered usability as an explicit goal of interactive systems. This work outlines and considers the barriers to developing differentially private interactive systems for informing public policy and offers an alternative way forward. We propose balancing three design considerations: privacy assurance, statistical utility, and system usability, we develop recommendations for making differentially private interactive systems work in practice, we present an example architecture based on these recommendations, and we provide an outline of how to conduct the necessary user-testing. Our work seeks to move the practical development of differentially private interactive systems forward to better aid public policy making and spark future research.
\end{abstract}

\section{Introduction}\label{sec:intro}
Access to administrative and survey data collected by federal agencies is essential for public policy research and evidence-based decision-making \citep{reamer2018counting}. However, privacy and confidentiality concerns often limit access to this valuable data \citep{drechsler2023differential, harris2017protecting}. Researchers typically face a choice between using limited public data or undergoing an extensive vetting process to access restricted data, which often also requires financial resources and traveling to metropolitan areas. This process is often prohibitive for some individuals, such as non-citizens, those from less-resources institutions, or individuals in geographically remote areas.

To address these barriers, the Committee on National Statistics \citep{national2024toward} and the Advisory Committee on Data for Evidence Building \citep{AdvisoryCommittee2022} recently recommended multi-tiered access frameworks. One such tier allows researchers to query targeted statistics for specific analyses or to validate public use data analyses. This tier offers higher-validity statistics to social scientists and policymakers without requiring full access clearance, streamlining the process and expanding accessibility.

We refer to this process as an interactive system, distinguishing it from a non-interactive approach where federal statistical agencies release pre-determined statistics or data files \citep{dwork2010differential}. This interactive method is the focus of our paper and is often referred to as a \textit{validation server} \citep{barrientos2021differentially}. We also focus specifically on the application of the differential privacy (DP) framework for privacy-preservation in interactive systems. DP, originally proposed by \cite{dwork2006calibrating}, and its surrounding framework has become one of the most studied privacy definitions in the statistical data privacy field, though there have been far fewer real-world applications which use DP \citep{panavas2024illuminating}. 

The system we describe would most closely resemble a structured query language (SQL) database\footnote{Other approaches have also been explored which allow more flexibility in the queries, e.g., \citet{tyagi2024privacy}.}, but the types of queries may be limited and noise would be added to the returned results to ensure differential privacy \citep{johnson2018towards}. The possible queries would include, for example, univariate or multivariate statistics, such as counts, means, quantiles, or linear regression estimates. This type of system allows for customization of queries to meet specific research questions, but, due to a variety of challenges, no government organization has currently implemented an interactive query system with formal privacy \citep{drechsler2023differential}. Even outside of government, at the time of writing this paper only LinkedIn has deployed a publicly available interactive query system \citep{rogers2021linkedin}.

This is largely because, in the interactive setting, certain incompatibilities exist between the DP framework and the standard statistical data analysis practice, which have created barriers to deployment of such DP systems for federal statistics. These issues have been detailed in \citet{snoke2024incompatabilities, drechsler2023differential,sarathy2023don}. We contribute to this discussion by reconsidering some of the underlying assumptions which have guided former work. While other papers start by assuming the DP framework and evaluating the feasibility of deploying a query system under that framework \citep{drechsler2023differential,barrientos2023feasibility, sarathy2023don}, we take the approach recommended in \citet{snoke2024incompatabilities} and attempt to find a \textit{compromise} between utilizing the ideas of DP and meeting the needs of the users. 

This paper may be best read as a call for a new research paradigm that challenges some of the assumptions of DP, while still seeking to maintain many aspects of the framework of DP. We recognize the significant value of the DP framework for formalizing notions of privacy, but we seek to merge this with the realities of statistical practice. In particular, we propose a system that allows data analysts to explore the data and receive valid statistical outputs without needing any knowledge of DP. In doing so, we articulate an alternative means of approaching the issue of interactive, privacy-preserving systems that focuses on usability. We do not claim that this approach is superior to other approaches in every way or without limitations. We instead claim that our approach seeks to find a solution to the problem in a novel way, which can better aid public policy making and spark future research. 

The remainder of the paper is organized as follows. In Section \ref{sec:dp_interactive}, we review the prior work on DP interactive query systems. In Section \ref{sec:design_consider},  we articulate design principles that a query system for federal statistics should achieve. In Section \ref{sec:barriers} we detail the issues these systems present which limit their ability to be used for informing public policy research. In Section \ref{sec:system_rec}, we provide proposed recommendations to include in the system and articulate how these recommendations seek a compromise between the principles of DP and the norms of statistical data analysis. In Section \ref{sec:infrastructure}, we outline a proposal of the infrastructure that could be used to meet these recommendations. Finally in Section \ref{sec:user_testing}, we discuss future user-centered research directions, and in Section \ref{sec:discussion} we provide a discussion of our framework and its future directions.

\section{Differential Privacy Interactive Query Systems}\label{sec:dp_interactive}
Differential privacy (DP) is a mathematical framework that provides a quantifiable and provable amount of privacy protection under given definitions of privacy loss \citep{dwork2006calibrating}. We do not provide an exhaustive review of DP definitions and methods. For further details from various perspectives, we recommend \cite{dwork2014algorithmic,wasserman2010statistical,snoke2020statisticians,slavkovic2023statistical} and \cite{wood2018differential}. In Section \ref{sec:design_consider}, we discuss the underlying concepts which form our definition of privacy assurance, which incorporates additional concepts beyond DPs.

Systems that rely on DP to provide access to confidential data are divided broadly into two main categories: non-interactive and interactive \citep{dwork2010differential}. Non-interactive systems generate and release the noisy results of predetermined queries, sometimes produced as microdata or summary statistics from the confidential data, aiming to address a range of analyses. Once the noisy statistics are released, they are final and in theory no other information can be obtained from the confidential data. This simplifies some of the decisions for the data providers, but it can create limitations if the choice of summary statistics or the amount of noise added to those statistics does not provide the information researchers need for their specific projects \citep{hawes2020implementing}.

In contrast, interactive systems (we will also refer to these as statistical query systems or validation servers) allow researchers and practitioners to directly query the data, providing them the flexibility to tailor their queries to their specific analyses. However, this flexibility introduces challenges in aligning with established DP protocols, as we will address in Section \ref{sec:barriers}. While a handful of non-interactive systems have been deployed (e.g., government agencies \citep{abowd2018us}, large tech \citep{aktay2020google, AppleDifferentialPrivacy2024}, non profits \citep{adeleye2023publishing}), interactive systems are much more rare. As mentioned previously, LinkedIn’s engagement API is the only publicly deployed interactive DP system. Other deployments are being explored (e.g., Internal Revenue Service) \citep{taylor2021privacy, tyagi2024privacy} or have been shut down \citep{near2018differential}.

Interactive DP systems are commonly provided in the context of an SQL database. This provides users with the experience of querying an SQL database, but the responses have noise added to protect privacy. Research in this area demonstrates that common SQL operations can be made DP, and organizations have developed libraries and infrastructure to help implement these solutions on customer databases \citep{near2018differential, wilson2019differentially, grislain2024qrlew}. However, while the technical aspects are well-understood, there has been little focus on user testing or gaining insights into whether this meets the practical needs of users. These systems assume that users are comfortable working in the SQL framework and can set their own privacy parameters \citep{grislain2024qrlew} or work with the designated accuracy provided \citep{OasisLabsPrivateSQL2024}.

\section{Design Considerations for Statistical Query Systems}\label{sec:design_consider}
Developing a DP statistical query system involves multiple stakeholders, such as data users, privacy specialists, and data stewards, each with different priorities and limited understanding of each other's domains. Painting with broad strokes, data users, who need data to conduct research or inform policy decisions, prioritize high-quality data for accurate and timely analyses. Privacy practitioners focus on applying rigorous methods to ensure privacy. Data stewards, often government officials, prioritize safely expanding access while reducing administrative burdens. While these groups may have overlapping priorities, each typically operates with a distinct set of criteria that are not always explicitly described \citep{cummings2023advancing, cummings2023centering}.

These conflicting goals are further complicated by differing interpretations of key terms like “high-quality data,” “privacy protections,” and “expanded access,” resulting in potential miscommunication and confusion. Without a shared understanding of success, it becomes difficult to design an effective system. To address this and provide a common language for further discussions in this paper, we establish three design considerations for private statistical query systems: \textit{privacy assurance}, \textit{statistical utility}, and \textit{system usability}. For each design consideration, we provide specific interpretations and outline criteria for effective implementation based on relevant literature and our own discussions with stakeholders. While others may use different criteria, or particular contexts may lead to even more detailed definitions, it is crucial to clearly articulate the chosen design considerations and definitions of success to ensure the system is both designed and evaluated effectively \citep{shackel2009usability}.


\subsection{Privacy Assurance}

When sharing government data, a patchwork of regulations attempt to ensure that no sensitive information about individuals is disclosed for unauthorized use \citep{biden2023executive, near2023guidelines}. In some cases, government-collected data, such as tax records, cannot be opted out of, and a breach in privacy could cause serious harm to individuals and damage governmental operations. Privacy assurance, as a design consideration, aims to minimize sensitive data leakage. There is (and likely always will be) a need for translation between laws and regulations and mathematical approaches for reducing privacy loss, since the regulations are not constructed with precise technical language applicable for every context \citep{seeman2023framing}. 

While this paper focuses on interactive DP systems, here we articulate three underlying tenants for evaluating the privacy assurance of a system that go beyond the framework of DP. We also provide them in language that is more interpretable for a broader audience. By doing so, we hope to articulate the practical goals of privacy assurance rather than couching them in technical formulations.

First, \textit{decision makers should be able to measure and track the privacy expenditure on individuals represented in the data over time}, such that it can inform decisions about future queries. The framework of DP offers one of the most tractable ways of performing privacy loss accounting, also known as composition \citep{mcsherry2009privacy,dwork2016concentrated,bun2016concentrated,kairouz2015composition}. Composition, broadly, enables a system to track the total privacy expenditure of a series of queries, taking into account the relationship between the individuals and the variables that inform the requested statistics.

Second, \textit{privacy assurance depends on whether we can transparently articulate the pieces of information that have been released without privacy protections and determine the additional risk they pose}. Any system will inevitably, in practice, release some information without rigorous privacy protections. Examples include the invariants (e.g., the unprotected state population totals) released in the 2020 Census \citep{abowd20222} and the Joint Committee on Taxation releasing tables based on administrative tax data outside of the IRS. More broadly, statistical analysis requires performing exploratory data analysis to understand the underlying data and determine the appropriate models. Removing EDA from statistical analysis workflows would represent a massive shift that the research field is not prepared to undergo, and currently DP does not provide the means of performing common EDA tasks \citep{snoke2024incompatabilities}. Some EDA or other data-driven tasks, such as hyperparameter tuning, may take place outside of the DP framework, but it should be articulated.

Finally, \textit{privacy assurance should include the ability to perform threat modeling to facilitate interpretations of privacy risks}. The DP privacy parameters alone do not provide interpretable guarantees of either the particular harms or the absolute risk of those harms to individuals \citep{slavkovic2023statistical}. Privacy guarantees are contextual and providing DP privacy parameters alone obscures some of the other potential harms \citep{seeman2024between}. A robust definition of privacy assurance includes the ability to define the potential harms, measure the empirical protections, and communicate this to decision-makers.

\subsection{Statistical Utility}
In an interactive system supporting public policy research, statistical utility should prioritize releasing data that enables making informed decisions \citep{hotz2022balancing}. The term utility is used broadly in the literature with many different meanings, so we need to carefully define it. We assess it along three key dimensions: \textit{(1) the scope of analyses that can be conducted under the privacy mechanism}, \textit{(2) the ability to make valid statistical inferences based on the available statistics}, and \textit{(3) the level of uncertainty induced by the privacy mechanism} \citep{williams2024disclosing}. 

Measuring statistical utility by the type of analyses that can be performed is important because it is easier to provide valid inference with less error on simpler statistics, like univariate statistics, under DP. In contrast, multivariate techniques, such as regression models, are more challenging to implement and often fail to produce outputs, such as confidence intervals, that correctly account for the uncertainty \citep{barrientos2023feasibility}.

When talking about the validity of estimates, we define the target of the inference as the true parameter of interest, rather than considering the distance between the noisy statistics and the confidential statistics. The latter does not necessarily help us answer policy questions because there is also uncertainty or measurement error in the confidential estimates. We separate validity from the level of noise added, since it is possible to have valid results with high levels of error or biased results with low levels of uncertainty. If the results are supplied with valid confidence intervals, statistical utility may still be degraded due to an increase in null findings (e.g., type II errors) over what would be expected with the confidential data.

\subsection{System Usability}
An effective statistical query system should be well designed to allow both data users and government agencies to easily adopt and use it. This concept may be implicitly grouped under the term ``statistical utility", but here we explicitly separate it in order to more rigorously define it and to give it equal weight as a design consideration. Usability, a less familiar term in statistics, refers to how easy and efficient it is for users (in this case meaning both system administrators and data users) to interact with a product, system, or service to achieve their goals. This depends on the users' required knowledge level, the design of the interface and documentation, and how well the system's outputs meet the specified user tasks and needs within various environments \citep{shackel2009usability}.

In designing DP systems, it is important to consider both parties: government agencies and data users. Government agencies provide the strategic direction and financial support, but without the active engagement of researchers, the system will not be effective or valuable. For government agencies, these considerations can be measured by infrastructure costs and administrative burdens. The decision processes about how to share data and apply statistical data privacy techniques should be clearly articulated, with straightforward guidelines in place. To help reduce staff burden, the system should be as automated as possible and should reflect existing data release procedures to increase the likelihood of adoption \citep{tyagi2024privacy}. For users, usability is measured by the ease of accessing the system and using it without extensive training. Requiring higher amounts of prior knowledge of privacy-enhancing technologies to access and interpret private data reduces usability \citep{sarathy2023don}.

Usability is commonly overlooked when designing DP systems \citep{cummings2023advancing, cummings2023centering}, and until workable solutions are created for these challenges, interactive systems will continue to be largely unusable and social scientists and policy makers will have reduced capabilities and accuracy when working with important government data. 

\section{Barriers to Interactive DP Systems}\label{sec:barriers}
The goal is to build an interactive DP system that balances the three design principles laid out in the previous section. However, meeting these design principles in practice reveals intricate challenges and trade-offs. Despite the substantial amount of theoretical work, a main reason that interactive DP systems have largely not been deployed is because of the numerous practical issues surrounding these systems \citep{seeman2024between}. These issues stem in part from incompatibilities between the DP framework and typical data science and analysis workflows, while others stem from unrealistic assumptions which privacy practitioners make about standard statistical analysts \citep{snoke2024incompatabilities, sarathy2023don}. In assessing these barriers, we focus particularly on areas where we see conflicts between the three design considerations.

\subsection{Exploratory Data Analysis}\label{subsec:eda}
One of the clearest ways that privacy assurance through DP fundamentally conflicts with traditional statistical analysis workflows is in the context of exploratory data analysis (EDA) \citep{nanayakkara2024measure}. On one hand, DP requires that any information which is derived from the confidential data must be perturbed and accounted for in the privacy budget \citep{amin2024practical}. On the other hand, good statistical practice assume a substantial amount of EDA. This creates a conflict between privacy assurance and statistical utility, since the privacy cost of any analyses would increase proportional to the amount of EDA. If the privacy cost is held fixed, performing more EDA would be expected to degrade the statistical utility of the final analysis due to increased uncertainty from the noise. Additionally, applying existing DP methodology limits the EDA techniques which can be used, since very little work exists on common methods such as data visualization, model selection, or diagnostic plots \citep{panavas2023investigating}.

Even if we assume data users know the exact query they wish to make without conducting EDA, as the prototype DP systems PSI \citep{gaboardi2016psi} and DPCreator \citep{sarathy2023don} assume, DP mechanisms require users to make assumptions about the domain and range of the confidential data and statistics they wish to estimate. This can produce conflicts with our other design principles. For example, users may not know how to make assumptions about the range of complex queries creating usability problems or understand how those different assumptions can significantly impact the statistical utility of the results \citep{williams2024benchmarking}.

Needing to make assumptions about the domain of the confidential data creates additional usability challenges. For example, if a researcher queries a regression model on a subset of the data with a coefficient for a category which has no observations in the subset, this will produce an error in standard OLS estimation procedures. But, deterministically returning this error violates the DP framework. The handling of errors should either be probabilistic—meaning errors are returned randomly, even for queries that don’t naturally produce them—or managed in a way that prevents their generation without relying on confidential data. This could be achieved by having accurate knowledge of the domain. We see this primarily as a usability challenge because a system that handles errors in an unfamiliar way will produce significant frustration and decrease adoption among potential users.

\subsection{Differential Privacy Parameter Setting}\label{subsec:dp_param}
Another conflict between our design principles stems from the need to select privacy parameters, e.g., $\epsilon$. In the original DP conception, the system administrators would determine the value of the privacy parameter (often called the ``privacy budget") which they are comfortable using and specify the value for each query. In practice, this is made more difficult by the fact that DP privacy parameters are difficult to interpret and select \citep{dwork2019differential, cummings2021need}. While attempts have been made at interpreting the privacy parameters \citep{wasserman2010statistical,kifer2022bayesian,near2023guidelines}), there remains no consensus in the field on an absolute standard. As \citet{slavkovic2023statistical} point out, this issue stems from the fact that DP measures the relative, rather than absolute privacy loss, so picking a privacy budget should be context dependent.

Other efforts \citep{tyagi2024privacy, grislain2024qrlew, aymon2024lomas} place the decision of setting the privacy parameters for each query in the hands of the data users, subject to an overall constraint. This creates challenges for the statistical utility of the system because the value of the privacy parameter directly impacts the amount of noise added to the outputs. Predicting the impact requires knowledge of the data generating distribution, which is often beyond the data users' capacity. The situation is analogous to a researcher preparing to field a survey without being able to conduct a power analysis to help determine the required sample size. 

In the few studies evaluating DP systems \citep{sarathy2023don, murtagh2018usable, ngong2024evaluating}, many users were either unable to set or were not confident in their parameters. Additional educational materials and visualization strategies have been used to help users set these parameters, but even these are not always successful \citep{nanayakkara2022visualizing, panavasvisualization}. 

\subsection{Fixed Privacy Budget}\label{subsec:fixed_budget}
The logic of DP dictates that the privacy budget should be a fixed value, but this concept creates particular issues in an interactive system. In non-interactive DP systems, data are released using a fixed privacy parameter and no additional privacy is expended regardless of how much users utilize the released data. However, in interactive systems, the privacy budget must not be capped, queries must only be run on new confidential data after a certain period, or ultimately the interactive system will become defunct. \citet{drechsler2023differential} note that the budget running out can cause a variety of issues, such as the ability to reproduce research results. Providing users with a system where they can run out of queries presents significant usability issues and assumes the need for extensive training. On the macro scale, government agencies are unlikely to want to build systems that will eventually become defunct.

A fixed privacy parameter also raises questions for statistical utility, since it must be efficiently allocated across many queries. In the interactive setting, all of the eventual queries may not be known in advance, so decision makers must allocate the privacy budget without being able to make fully informed decisions. This may result in some queries consuming a disproportionate share of the budget, either using more privacy than necessary for meaningful statistical analysis or limiting the utility of the outputs where it is needed. Setting a fixed value across different users assumes that each analysis plan has the same complexity and importance and that all study populations are of the same size. However, this is not generally true in practice \citep{drechsler2023differential}.

\subsection{Private Result Interpretation}\label{subsec:private_inference}
Finally, the DP literature assumes that users who receive noisy statistics will understand how to properly analyze them and report on the impact of noise infusion for their analyses. This is unlikely to be true in practice without extensive training or documentation, leading to a usability challenge. For example, \citet{boyd2022Differential} noted that when DP statistics were released for the 2020 Decennial Census data product series, people unfamiliar with how to interpret the noise became distrustful of the data as a whole. Similarly, researchers who were interviewed about using DP systems worried specifically about how to explain the results when publishing their findings \citep{sarathy2023don}.

Users unfamiliar with DP will likely not know how to generate valid confidence intervals unless explicitly provided, or they might not know how to appropriately use the noisy results for other downstream analyses. More fundamentally, even if the statistics are properly analyzed, users would likely struggle to adequately explain the impact of noise infusion to to reviewers, collaborators, or any audiences assessing the study's results.  Federal agencies face related challenges when determining how to report noisy statistics to different users who may request overlapping queries \citep{snoke2024incompatabilities}. There are currently no standard practices for when or how to report the level of noise added, e.g., the privacy parameter or the assumptions underlying how the confidence intervals were generated, in a way that can be easily interpreted by common statistical users \citep{wood2018differential, oberski2020differential}.

\section{System Recommendations}\label{sec:system_rec}
In our recommendations, we prioritize system usability while balancing privacy assurance and statistical utility. DP research has traditionally focused on strict privacy guarantees, occasionally considering statistical inference but rarely usability \citep{sarathy2023don, ngong2024evaluating}. We aim to shift this focus, recognizing that without usability, systems are unlikely to be adopted. As DP continues to move from theory to practice, our goal is to enhance usability with minimal compromises to privacy and statistical utility, ensuring systems can be both practical and effective.

Based on this weighting of the design criteria and the articulated challenges related to interactive DP systems, we propose several system recommendations for government organizations attempting to implement an interactive query systems. For each one, we consider how it impacts the three design considerations and explain why the benefits may outweigh the limitations. We aim to provide a holistic approach that demonstrates how usability can be enhanced without significantly compromising privacy or statistical utility. A summary of the recommendations is provided in Table \ref{tab:recommendations}.

\subsection{Synthetic Data for Exploration}\label{subsec:synth_eda}
For an interactive query system to be successful, we propose that there must be synthetic data or a public file that resembles the original data available as part of the system \citep{sarathy2023don}. Users can explore the synthetic data to understand the shape and structure of the variables, devise queries which will not produce errors, and refine their understanding of how the privacy mechanisms will impact their analyses. For the synthetic data to work, it needs to contain, at a minimum, the same metadata, the same data domain, and at least some similarity in the statistical outputs range. 



\textit{Privacy Assurance:} Providing synthetic data should improve usability and statistical utility, but it may reduce privacy assurances. Creating synthetic data can compromise privacy assurances, as some privacy is inevitably leaked by producing a dataset that conditions on the real data. This could be mitigated by using DP methods to generate the synthetic data, but this may not be feasible in every situation \citep{jordon2022synthetic}. In another way, providing synthetic data may mitigate privacy risks by limiting the need for repeated queries on the real dataset. Without an exploratory data set, users might inadvertently increase privacy risks by generating ineffective queries, leading to more frequent system interactions and a greater potential for privacy loss. 

Permitting EDA interactively will likely result in a high privacy budget usage. It may be more effective to apply that cumulative privacy loss once to provide users with an understanding of the data in a synthetic data format. Although the use of synthetic data might reduce privacy assurance, this risk might also be managed by ensuring that the synthetic data is not directly based on the same confidential data, such as by using datasets from prior years \citep{el2020practical}.

\textit{Statistical Utility:} Allowing users to explore the data and experiment with different queries can help them better determine the types of queries needed for their analysis. Incorporating this process into their data analysis workflows will also give them a clearer understanding of the impact of choosing different privacy parameter values on the uncertainty in their analyses. The challenge for statistical utility would be determining what relationships to preserve in the synthetic data. For example, \citet{aymon2024lomas} offer a solution for using synthetic data generated from public metadata to test queries. But, because the data is randomly generated, it does not provide users with anything beyond getting the syntax correct. 

Synthetic data would ideally preserve at least simple marginals and the approximate range of the variables to give an accurate sense of the data for common queries. However, users need to understand that results from confidential data may differ significantly. For example, null findings on synthetic data could discourage users from pursuing analyses on the confidential data, potentially leading to missed insights.

\textit{System Usability:} Synthetic data helps enable users to conduct exploratory data analysis \citep{el2020practical}. With synthetic data, users could experiment with query syntax and outputs, ensuring they do not submit invalid or uninformative queries \citep{aymon2024lomas}. This approach is similar to other statistical workflows, where researchers conduct pilot studies or use smaller samples before a full-scale analysis, thus preserving familiar processes for public policy research. By using synthetic data, they can help ensure the query results they obtain will work for their downstream analysis. Importantly, they could explore the data using tools and processes they are familiar with before having to make more constrained queries. A challenge in terms of usability would be the administrative burden of creating the synthetic data \citep{bowen2020synthetic, bowen2022synthetic}.

\subsection{Remove Privacy Parameter Language}\label{subsec:remove_epsilon}
We propose to avoid the need for users to interpret and set the differential privacy implementation parameters (e.g., privacy parameters ($\epsilon$, $\delta$),  contextual parameters (data bounds)) by removing them from the user inputs. Even with documentation provided, individuals unfamiliar with differential privacy struggle to understand the implementation choices and parameters \citep{murtagh2018usable}. Instead, we propose that users request a desired level of expected statistical accuracy, e.g., in the form of bounds on the expected error or in the form of additional uncertainty \citep{ligett2017accuracy,williams2024disclosing}. This would then be automatically translated into an appropriate privacy parameter value to be evaluated by the data administrators (more on this in Section \ref{subsec:human_review}), removing the need for users to understand or set implementation parameters themselves. To broaden the scope of supported queries and reduce user confusion, data curators could add key metadata parameters to contextualize the data \citep{aymon2024lomas, sarathy2023don}. For simplicity, any reference to implementation parameters aside from accuracy should be omitted from the user interface.



\textit{Privacy Assurance:} The privacy assurance depends on the ability to compute the privacy parameters based on the user inputs without using the confidential data. Where this is possible, the process we recommend will not degrade the privacy assurance. An alternative approach could utilize information about the confidential data (i.e., estimates of the data generating distribution or the uncertainty inherent in the confidential statistics) to more precisely compute the privacy parameters. In this case, the value of the privacy parameters could not be released without creating additional privacy loss. Further, the data administrators still need to rationalize about whether the requested accuracy corresponds to too much privacy loss. They may feel pressure to approve high values of privacy parameters to satisfy users' needs.

\textit{Statistical Utility:} Removing the privacy parameter enables users to make queries with an appropriate level of accuracy that supports their analysis, enhancing statistical utility. In some cases, their request may not be feasible, but this approach reduces the possibility that a user unknowingly requests a statistic that has too much noise added to be of use for their analysis. The other potential drawback is that in some cases there are multiple accuracy metrics which users may care about, and it may not be clear how to select the right one. While some solutions have been proposed, see \citep{ligett2017accuracy, rogers2024adaptive}, there is still methodological work that needs to be done in the field to ensure accurate transformations from the utility metrics to the privacy parameters is possible. In some situations, it may require simulations to map expected accuracy to privacy parameters, and truncating values may further complicate this mapping.

\textit{System Usability:} Not requiring users to make determinations about an unfamiliar parameter, which is outside the purposes of their analyses, should increase the usability of the system. Statistical researchers and data practitioners are familiar with reasoning about metrics such as bias, standard errors, or statistical power, so we recommend focusing on this language in the user interface. This approach aligns better with the existing expertise of data users, and it simplifies the task for system administrators who no longer need to provide extensive documentation about selecting implementation parameters. The system administrators will have to take on the extra task of deciding on contextual parameters, namely bounds, but will be far more equipped to do so than data users. Additionally, as we will discuss in the next recommendation, asking users directly about their requirements on the accuracy and then translating this into a corresponding level of privacy loss will help streamline the approval process. This process shares some similarities with selecting a sample size in experimental or survey design, which should be familiar for some data users.

\subsection{Per-Research Question Privacy Allocation}\label{subsec:per_question_eps}
Setting the appropriate privacy budget is a socio-technical question that should be based on balancing the tradeoffs between user harm and statistical utility \citep{sarathy2023bridging}. We propose using a per-research proposal budget to ensure privacy is allocated efficiently. Existing systems \citep{rogers2021linkedin} assume a per-user privacy budget and give each user the same fixed privacy budget. Instead, users would prepare to submit a set of queries for their project and request a specific level of accuracy to determine the total privacy budget for that research proposal.


\textit{Privacy Assurance:} This type of system has no cap on the total privacy budget, so system administrators would need to carefully consider each additional request to determine whether it is appropriate \citep{rogers2016privacy}. Ultimately, we believe this approach does not worsen the problem compared to existing methods for managing per-user privacy budgets, which are usually large and periodically refreshed \citep{rogers2021linkedin}. Using per-project allocations does not guarantee more privacy loss, it simply shifts the source of the allocation. One limitation is that privacy budgets may not be given equally to different users or researchers, which raises some questions of equity. A per-project privacy budget, however, only uses the privacy budget on specifically requested queries and aligns the privacy allocation more closely with the intended uses.

\textit{Statistical Utility:} Similar to usability, a per-project allocation would enable researchers to design their queries without the constraint of a single privacy budget, allowing them to consider the necessary accuracy for each query. On the other hand, for particularly complex projects it may be difficult to accurately estimate the required privacy budget to meet all of the accuracy requirements specified by the users. Because that process will entail approximations, it is possible that on a per-project level the approximations become poor and the desired level of accuracy is not achieved.

\textit{System Usability:} This approach enhances system usability for both data users and administrators. First, the system mirrors existing processes for accessing restricted data, where researchers must carefully examine and justify their need for specific datasets and levels of data granularity. It also shares similarities with common practice at institutional review boards (IRBs), which must balance the potential harms of a research study against the potential benefits to society. Data users do not traditionally have to work within constraints of a predetermined privacy budget which may not fit the complexity of their project. 

This also pairs well with removing the burden of correctly allocating a fixed budget across a set of queries, discussed in Section \ref{subsec:fixed_budget}. Additionally, system administrators would be relieved from the burden of predetermining a fixed privacy allocation that would work for a wide range of projects. In reality, projects seeking to use the federal statistical data vary broadly in the number and type of statistics they require. Additionally, neither administrators nor users need to worry about what happens when a privacy budget is exhausted, since researchers could submit future proposals for follow-up work if necessary \citep{drechsler2023differential}.

\subsection{Light Human Review}\label{subsec:human_review}
In light of the preceding two recommendations, we recommend maintaining some level of human review in the system. When submitting a proposal, users would include several paragraphs explaining the purpose of their research and justifying the requested level of accuracy. Submitting analogous forms is common practice for researchers who have previously requested government data or submitted to an IRB. The trained individual would review the automatically compiled report consisting of the requested queries, project justification, required accuracy, proposed work products, and computed privacy parameters. This review would occur before the queries are approved and run on the confidential data. The review can be approved, denied, or it can be sent back with adjusted levels of accuracy if the requested levels incurred too much privacy loss. The data user can consider whether to accept the adjusted levels or to reject them if they decide it will not meet their analysis needs.

\textit{Privacy Assurance:} A human reviewer might make it more difficult for a system user to conduct privacy attacks that could otherwise bypass automated checks. Conversely, as we mentioned previously, incorporating human discretion could lead to increased privacy loss or inequitable distribution of privacy loss due to the desire to meet users' needs. Additionally, if the report includes information errors generated by the proposed queries (e.g., due to empty subsets), the reviewer can reject the proposal, but they would need to decide whether to disclose the reasons. If they did, that could lead to additional privacy leakage, but if we do not it might lead to issues of trust in the system.

\textit{Statistical Utility:} Human review can help avoid data users and researchers submitting queries which return errors. While this process would not be expected to change anything about the way results are returned, the human-in-the-loop may help ensure that researchers conduct their analyses more accurately. 

\textit{System Usability:} This process improves usability by enabling the two prior recommendations, as well as by avoiding issues that currently have no solutions, such as error handling. The review process can be designed to be simple and quick, since the necessary information can be distilled into a summary report. This process would add both user and administrative burden over a fully automated system, but a human review process can help reduce time to deployment. Participants in the study by \citet{sarathy2023don} specifically requested a similar process to increase their comfort with publicly releasing private data. Future methodological work might further reduce the need for human review, but currently maintaining some level of review would enable agencies to create interactive systems faster.

\subsection{Output Documentation:}\label{subsec:release_language}
Finally, we recommend presenting results of the queries in a familiar manner for users, similar to what they would see if the data were not subject to DP. This presentation should include automatically calculated uncertainty measures that incorporate DP noise, enabling users unfamiliar with DP to confidently integrate the data into their analyses. This also means that the system can only incorporate DP mechanisms, which enable the release of uncertainty estimates. This is not true of the majority of work in the field \citep{barrientos2023feasibility}. Additionally, to further support researchers, we propose automatically generating example language about the validity of the outputs that can be directly included in academic publications. These paragraphs will clearly communicate how the results were generated, estimates of the level of noise introduced, and how valid inferences can be made.

\textit{Privacy Assurance:} If the language used is standard and not based on the confidential data, this recommendation should not impact the privacy guarantees of the system.

\textit{Statistical Utility:} Including standard language and documentation will help data users and researchers appropriately analyze the results of the queries. This process will also help the broader field grow in familiarity with these type of noisy estimates and how we should handle them, similar to other common statistical issues such as missing data or measurement errors. These best practices will increase the ability to perform research with new data sources.

\textit{System Usability:} Formatting the data release in a language and structure familiar to users will simplify their ability to analyze the data and publish results. A significant challenge identified in several studies on DP usability is the difficulty users encounter in interpreting, using, and discussing private data outputs \citep{cummings2023centering, sarathy2023don, cummings2023advancing, boyd2022Differential}. Determining the necessary code and drafting the academic language will require an investment of effort, but it will prevent users from having to guess how to appropriately present their results each time and should be easily replicable.
\begin{table}[h!]
\centering
\caption{Summary of Design Considerations for System Recommendations}
\label{tab:recommendations}

\renewcommand{\arraystretch}{1} 

\begin{adjustbox}{max width=\textwidth}
\begin{tabular}{|p{4cm}|p{4cm}|p{4cm}|p{4cm}|} 
\hline
\textbf{Recommendation} & \textbf{System Usability} & \textbf{Statistical Utility} & \textbf{Privacy Assurance} \\ \hline

\textbf{Synthetic Data for Exploration} (Section \ref{subsec:synth_eda}) & 
\RaggedRight Allows users to explore and test queries. Some admin burden in generating synthetic data. & 
\RaggedRight Improves understanding of privacy's impact on analyses. & 
\RaggedRight May leak some privacy; mitigated by differentially private methods. \\ \hline

\textbf{Remove Privacy Parameter Language} (Section \ref{subsec:remove_epsilon})& 
\RaggedRight Simplifies interface by eliminating privacy parameters. Focuses on accuracy familiar to users. & 
\RaggedRight Enables more accurate queries. Challenges arise if multiple accuracy metrics are needed. & 
\RaggedRight Doesn't affect privacy if privacy parameters are pre-computed. Risk increases if based on real data. \\ \hline

\textbf{Per-Research Question Privacy Allocation}  (Section \ref{subsec:per_question_eps}) & 
\RaggedRight More flexibility, reducing admin/user burden by allocating budget per project. & 
\RaggedRight Enables better query design without fixed budget constraints. & 
\RaggedRight No fixed cap on privacy budget; requires careful review to maintain privacy. \\ \hline

\textbf{Light Human Review} (Section \ref{subsec:human_review}) & 
\RaggedRight Increases user comfort with query vetting; adds admin time. & 
\RaggedRight No direct impact, but reduces user errors. & 
\RaggedRight Prevents privacy attacks that bypass automated systems. \\ \hline

\textbf{Output Documentation} (Section \ref{subsec:release_language}) & 
\RaggedRight Formats data releases in familiar language; auto-generates publication-ready text. & 
\RaggedRight Improves users' ability to analyze and report noisy estimates. & 
\RaggedRight No privacy impact if output language is standard and generic. \\ \hline

\end{tabular}
\end{adjustbox}
\end{table}

\section{Infrastructure Proposal}\label{sec:infrastructure}
Based on the recommendations outlined, we propose a high-level schematic of a private interactive query system infrastructure focused on usability and the federal statistical system. While only a theoretical system, we demonstrate how our design considerations and recommendations can guide the creation of a functional system. Some questions remain unanswered and will require organization-specific discussions, but this proposal offers a concrete and actionable framework to assess the feasibility of our recommendations. Additionally, our proposed infrastructure identifies gaps in understanding or best practices. We use these to suggest research directions aimed at addressing the challenges of making an interactive system fully operational in Section \ref{sec:user_testing}.

To support implementation, we strongly recommend using tested, well-documented, and actively maintained open-source differential privacy libraries whenever possible \citep{Shoemate_OpenDP_Library, berghel2022tumult}. These libraries enable faster deployment, offer rigorously tested differential privacy functions, and enhance reproducibility for other researchers \citep{taylor2021privacy}. If additional functionality is needed beyond what these libraries provide, researchers should ensure that all code is open-sourced to promote reproducibility and allow for independent verification \citep{casacuberta2022widespread}.

\begin{figure}[!h]
    \centering
    \includegraphics[width=\textwidth]{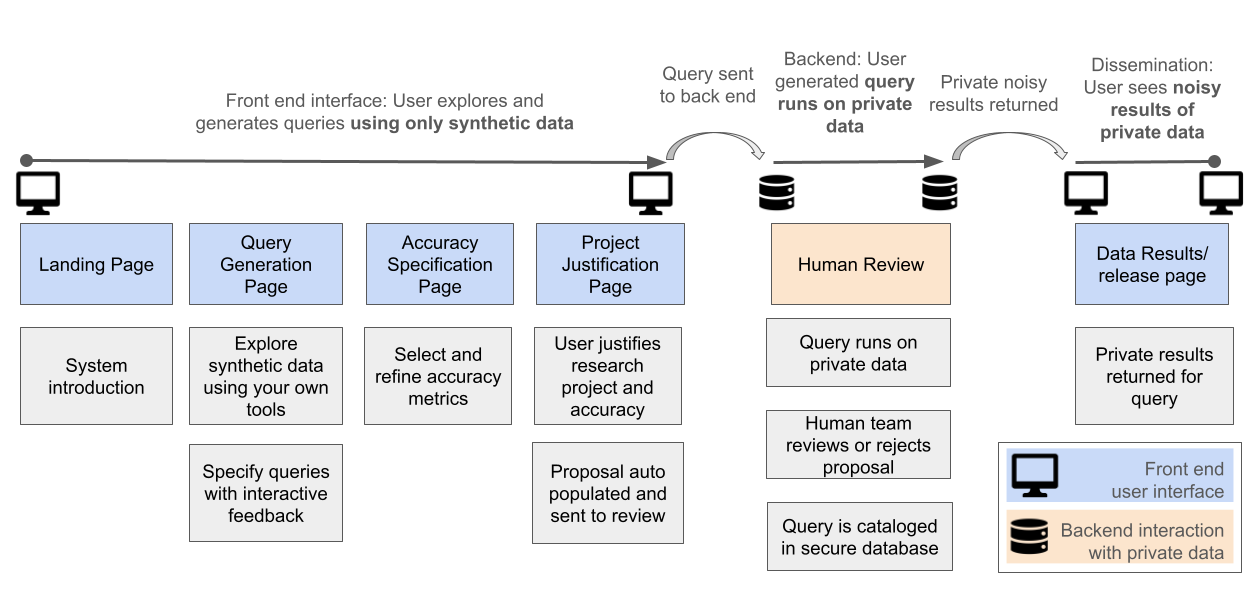}  
    \caption{\textit{This diagram provides a step-by-step overview of the interactive query system. The process begins with users exploring and generating queries using only synthetic data on the front-end interface (blue boxes). Users then justify their project and privacy parameters before their queries are sent for human review (orange box). The queries are run on the confidential data, and if no errors are found and query is approved the noisy results are returned to users (final blue box).}}
    \label{fig:count_validation_server}
\end{figure}

Figure \ref{fig:count_validation_server} shows an overview that can be read from left to right and top to bottom illustrating the entire process. 
To illustrate how this proposed infrastructure would work, we walk through an example of how a researcher or data user would use our system. Emily, is a public policy researcher and statistician working at a large non-profit. She is investigating whether Black and Hispanic families are claiming the Child Tax Credit (CTC) at lower rates than White families and wants to use IRS data on CTC claims stratified by race. She hopes to use her findings to have informed targeted outreach and policy interventions that ensure equitable uptake of the tax credit.

After learning about the interactive query system run by the IRS\footnote{Recall, this is an illustrative example of a theoretical system.}, Emily goes to the website and reads through the \textit{landing page}. She learns that there are synthetic data she can use, but that it should not be used for statistical inference. Emily then goes to the \textit{query generation page} where she downloads the synthetic data and imports it into her regular working environment. She uses the synthetic data to find out about how many records are available for each racial group for her analysis. After getting a sense of the quantity of records she generates her queries using an interactive user interface that helps systematize the query generation process and ensure there are no syntax errors. A preview output is generated using the synthetic data to ensure Emily requests the variables she needs for her analysis.

Once Emily finalizes the data request, she proceeds to the \textit{accuracy requirement page}. Here Emily inputs her accuracy requirements for each of the queries. A table displaying what those outputs would look like if generated from the synthetic data is shown to help Emily decide if the accuracy is appropriate. Once she finalizes her accuracy needs, Emily fills out a short form describing her research project, analysis goals, and accuracy requirements. She clicks submit and her data request and responses are sent to the government administrator for review. 

Once the data request is sent, an algorithm determines the privacy loss parameter required for each query. A report containing the queries, any errors they cause, privacy loss parameters, planned output, and the user's justification is securely sent to a trained reviewer in the government agency for \textit{human review}. The reviewer then goes through guidelines to determine if the requested data should be released to Emily. The reviewer can approve, reject, or inform Emily that certain queries need accuracy relaxations. Once the request is approved, Emily can see the results of the queries. The \textit{data release page} contains a table with the private outputs to download and example language for Emily to use if she publishes work related to this data.

\section{User Research}\label{sec:user_testing}
There are many details and components in our proposed infrastructure that require user testing or additional research to implement successfully. Many challenges and research questions arise regardless of the system put in place. However, by examining this proposal critically, we focus on key areas where further research is required and offer ways of initiating this research.


We propose integrating insights from human-computer interaction (HCI) research to study system usability—specifically, how individuals interact with DP systems. HCI is a discipline focused on understanding how people interact with technology and designing systems that are user-friendly and effective. In the context of DP, applying HCI principles allows us to evaluate not only whether our proposed recommendations improve system usability but also how best to implement them in real-world settings. By conducting usability studies \citep{sarathy2023don}, we can assess user preferences \citep{cummings2021need}, measure the practicality of our recommendations \citep{dumas2007usability}, and determine the most effective ways to integrate these systems \citep{garrido2022lessons}.

Below, we outline several research questions that are applicable to our proposal and could broadly impact a variety of differential privacy systems. We briefly outline the reason for the question and discuss which HCI methodologies would be applicable in solving these challenges. While there are many open research questions in relation to privacy assurance and statistical utility, we hope to provide a starting point for researchers interested in exploring \textit{system usability} challenges in DP systems. A summary of the questions, along with the corresponding barriers they seek to address and recommendations they seek to fulfill, is provided in Table \ref{tab:barriers_recommendations}.

\noindent \textbf{Q1: Does adding synthetic data alongside a validation server improve users' ability to generate private queries for analysis?}

A key issue in DP systems is the limited capacity for exploratory data analysis \citep{sarathy2023don, nanayakkara2024measure}. One potential solution, seen in both our recommendations and other interactive query systems, is the inclusion of synthetic data to restore a more familiar data science workflow. However, further studies are needed to explore whether synthetic data actually enhances user performance, and how similar to the confidential data the synthetic data needs to be for it to be effective. Such studies could compare user outcomes across systems with no synthetic data, randomly generated synthetic data \citep{aymon2024lomas}, and synthetic data produced by state-of-the-art algorithms \citep{bowen2021comparative}. User performance could be evaluated through metrics such as efficiency (time taken to complete analyses), accuracy (ability to make correct statistical inferences), and usability (cognitive load measured by the NASA-TLX scale) \citep{hart2006nasa}. Understanding the potential benefits of synthetic data could help organizations optimize both user processes and resource allocation.

\noindent \textbf{Q2: Which accuracy metrics would users prefer to set for their requested queries, and how well can they interpret the impact of these requirements on downstream analysis tasks?}

Literature on DP shows that future systems could ask for accuracy first requirements \citep{ligett2017accuracy}, but the types of accuracy metrics used and accepted are varied \citep{panavas2024illuminating}. Aside from the standard metrics chosen to evaluate algorithms in DP literature, we know very little about which data utility metrics people would prefer to see or use to evaluate the quality of their data. Researchers could conduct interviews with end users of DP deployments and use a deductive qualitative analysis to map the types of accuracy metrics used for different types of analysis \citep{azungah2018qualitative}. Quantitative user studies could examine users' decision-making processes and downstream analysis accuracy as they set accuracy thresholds for hypothetical analysis \citep{nanayakkara2024measure}. Results from this study could inform systems beyond interactive query systems as researchers could present their empirical evaluations of their algorithms in an end user context, provide better benchmarking results, and inform privacy engineers how to communicate the accuracy of their noisy data \citep{williams2024benchmarking}.

\noindent \textbf{Q3: What processes and guidelines ensure effective and fair evaluation of research proposals requiring private queries?}

The challenge in developing guidelines for approving certain privacy budgets arises from the inherent complexity of balancing privacy with statistical utility and the absence of universally accepted criteria for decision-making \citep{cummings2024attaxonomy, miklau2022negotiating}. Without a consensus on the optimal privacy parameter value in every situation, establishing a formal evaluation process is essential. Rather than setting fixed numerical thresholds, we propose exploring a structured, consensus-driven approach, leveraging expert reviewers to ensure decisions are both repeatable and consistent.

The general process would begin with assembling a corpus of research proposals, each specifying a requested accuracy level and corresponding privacy parameter. A panel of experts would develop an initial set of guidelines and evaluative questions to determine proposal approval. Experts would independently rate the proposals, then convene to discuss and address discrepancies, refining the guidelines based on their findings. This iterative process would continue with subsequent batches of proposals until consistency in decision-making is achieved, measured using inter-rater reliability (IRR). Once an IRR of 0.8 is reached, the guidelines can be considered robust for consistent evaluations \citep{hallgren2012computing}.
Through this structured, consensus-driven approach, the guidelines evolve to provide repeatable and reliable evaluations, offering a practical starting point for organizations assessing DP releases.

\noindent \textbf{Q4: What guidelines and documentation are necessary to ensure the accurate and consistent reporting of statistical analyses derived from differentially private methods?}

Previous user studies have shown that researchers often struggle with understanding how to use and publish results derived from differentially private statistics, such as for peer-reviewed articles \citep{sarathy2023don}. To address this, our recommendation in Section \ref{subsec:release_language} encourages federal agencies to provide data users with clear text, documentation, and guidance for publishing statistical analyses derived from differentially private results. While this recommendation is crucial, it remains an unresolved issue requiring input from various stakeholders, including privacy researchers, publishers, and social scientists. 

Other fields, such as medical research (e.g., CONSORT \citep{schulz2010consort}, STROBE \citep{vandenbroucke2007strengthening}) and machine learning (e.g., TRIPOD \citep{moons2015transparent}), have faced similar challenges with disorganized and incomplete reporting guidelines. To address these issues, they have successfully developed widely adopted reporting standards. The medical field, in particular, offers a systematic approach to creating guidelines that could be adapted to differential privacy \citep{moher2010guidance}. Their methodology typically involves three stages: defining initial criteria, convening diverse stakeholders face-to-face to deliberate, and producing a final checklist that outlines essential items for reporting. This checklist often includes clear reporting requirements and example text, serving as a standard for evaluating study results. 

A similar process could be used to develop comprehensive guidelines for reporting results in differential privacy. Federal agencies could use these guidelines to provide example analysis and results templates to help users include the necessary materials and language in their publications \citep{hoffmann2014better, yu2021automated}. These guidelines could include recommendations for organizations releasing private data to explain critical caveats, such as variability in responses to identical private query requests from different researchers. The same process could also address another significant gap: creating guidelines for publishing differentially private statistics in non-interactive settings \citep{dwork2019differential}

\begin{table}[h!]
\centering
\caption{Barriers, Recommendations, and User Research for Interactive DP Systems}
\label{tab:barriers_recommendations}
\begin{adjustbox}{max width=\textwidth}
\begin{tabular}{|p{4cm}|p{5cm}|p{5cm}|}
\hline
\textbf{Barriers} & \textbf{System Recommendations} & \textbf{User Research} \\ \hline

\parbox[t]{4cm}{\raggedright Exploratory Data Analysis (Section \ref{subsec:eda})} & 
\parbox[t]{5cm}{\raggedright Synthetic Data for EDA (Section \ref{subsec:synth_eda})} & 
\parbox[t]{5cm}{\raggedright Q1: Does adding synthetic data alongside a validation server improve users' ability to generate private queries for analysis?} \\ \hline

\parbox[t]{4cm}{\raggedright Differential Privacy Parameter Setting (Section \ref{subsec:dp_param})} & 
\parbox[t]{5cm}{\raggedright Remove Privacy Parameter Language (Section \ref{subsec:remove_epsilon})} & 
\parbox[t]{5cm}{\raggedright Q2: Which accuracy metrics would users prefer to set for their requested queries, and how well can they interpret the impact of these requirements on downstream analysis tasks?} \\ \hline

\parbox[t]{4cm}{\raggedright Fixed Privacy Budget (Section \ref{subsec:fixed_budget})} & 
\parbox[t]{5cm}{\raggedright Per-Research Question Privacy Allocation (Section \ref{subsec:per_question_eps}) \newline Light Human Review (Section \ref{subsec:human_review})} & 
\parbox[t]{5cm}{\raggedright Q3: What processes and guidelines ensure effective and fair evaluation of research proposals requiring private queries, fostering consensus among reviewers?} \\ \hline

\parbox[t]{4cm}{\raggedright Private Result Interpretation (Section \ref{subsec:private_inference})} & 
\parbox[t]{5cm}{\raggedright Output Documentation (Section \ref{subsec:release_language})} & 
\parbox[t]{5cm}{\raggedright Q4: What guidelines and documentation are necessary to ensure the accurate and consistent reporting of statistical analyses derived from differentially private methods?} \\ \hline

\end{tabular}
\end{adjustbox}
\end{table}

\section{Discussion}\label{sec:discussion}
This paper establishes a framework for implementing DP in a way that promotes collaboration between key stakeholders, such as privacy specialists, researchers, and policymakers. This framework is designed to encourage ongoing dialogue and joint decision-making around system design, ensuring that all parties' needs are considered. Additionally, while theoretical models may not always offer concrete guidance on practical compromises, empirical and user-driven guidelines rooted in HCI principles can help navigate these uncertainties. Other fields, such as machine learning, have successfully integrated similar approaches, improving the usability and transparency of their systems \citep{amershi2019guidelines}.

While DP is promising in its ability to protect individual data, it presents significant challenges when applied in practice to statistical analyses. For statisticians, social scientists, and public policy researchers, the strict constraints of DP often make data less accessible and harder to analyze effectively. To make DP systems functional in practice, practical modifications must be considered. Such modifications could balance the need for privacy with the requirements of statistical analysis, ensuring that these systems are both secure and usable.

Theoretical formulations of DP provide a well-defined yet narrow mathematical framework to assess privacy and utility. However, when implementing these systems in practical settings, the tradeoffs become far less clear as a wide range of social technical issues come into play. Effective solutions will require the involvement of government officials, statisticians, social scientists, and policymakers early in the design process, as their expertise is essential in ensuring that privacy systems remain functional for end users. Unfortunately, there is limited research on how users, particularly these various stakeholders, interact with these systems. This paper provides a framework for defining the desired design principles of a practical interactive system, provides recommendations to meet these principles, and outlines methods for using HCI research concepts to test out infrastructure proposals. While only a first step, the ideas laid out in this paper suggest a way forward for creating DP interactive systems that can be used.

\section*{Acknowledgments}
\noindent This research was funded by the National Science Foundation National Center for Science and Engineering Statistics [49100422C0008].

We would like to thank our collaborators Andr\'{e}s Felipe Barrientos, Leonard Burman, Graham MacDonald, Rob McClelland, Deena Tamaroff, and Silke Taylor on the Safe Data Technologies Project team for fruitful conversations and input on this work.

\subsection*{Contributions}
We used the CRediT taxonomy\footnote{See website to learn more about the CRediT taxonomy, \url{https://credit.niso.org}} to indicate author contributions to this paper.
\noindent LP: Conceptualization, Investigation, Methodology, Visualization, Writing – original draft, Writing – review \& editing \\
\noindent JS: Conceptualization, Investigation, Methodology, Writing – original draft, Writing – review \& editing \\
\noindent ET: Conceptualization, Investigation, Writing – review \& editing \\
\noindent CMB: Conceptualization, Funding acquisition, Investigation, Project administration, Writing – review \& editing\\
\noindent ARW: Conceptualization, Investigation, Writing – review \& editing
 
\section*{Disclosure Statement}
The authors report there are no competing interests to declare.

\bibliographystyle{chicago}
\bibliography{Bibliography-MM-MC}

\end{document}